\documentclass[%
 preprint,
 amsmath,amssymb,
 aps,
]{revtex4-1}

\usepackage{graphicx}
\usepackage{dcolumn,color}
\usepackage{bm}
\usepackage[mathlines]{lineno}

\begin{document}

\preprint{CHAOS/FARANDA}

\title{Interrupting vaccination policies can greatly spread SARS-CoV-2 and enhance mortality from COVID-19 disease: the AstraZeneca case for France and Italy}

\author{Davide Faranda}
\email{davide.faranda@cea.fr}
\affiliation{Laboratoire des Sciences du Climat et de l’Environnement, CEA Saclay l’Orme des Merisiers, UMR 8212 CEA-CNRS-UVSQ, Université Paris-Saclay \& IPSL, 91191, Gif-sur-Yvette, France.}
\affiliation{London Mathematical Laboratory, 8 Margravine Gardens, London, W6 8RH, UK.}
\affiliation{LMD/IPSL, Ecole Normale Superieure, PSL research University, 75005, Paris, France.}

\author{Tommaso Alberti}
\affiliation{National Institute for Astrophysics-Institute for Space Astrophysics and Planetology (INAF-IAPS), via del Fosso del Cavaliere 100, 00133, Rome, Italy.}

\author{Maxence Arutkin}
\affiliation{UMR CNRS 7083 Gulliver, ESPCI Paris, 10 rue Vauquelin, 75005 Paris, France.}

\author{Valerio Lembo}
\affiliation{National Research Council of Italy, Institute of Atmospheric Sciences and Climate (CNR-ISAC), Bologna, Italy.}

\author{Valerio Lucarini}
\affiliation{Department of Mathematics and Statistics, University of Reading, Reading, UK.}
\affiliation{Centre for the Mathematics of Planet Earth, University of Reading, Reading, UK.}
    
\date{\today}

\begin{abstract}
Several European countries have suspended the inoculation of the AstraZeneca vaccine out of suspicion of causing deep vein thrombosis. In this letter we report some Fermi estimates performed using a stochastic model aimed at making a risk-benefit analysis of the interruption of the delivery of the AstraZeneca vaccine in France and Italy. Our results clearly show that excess deaths due to the interruption of the vaccination campaign injections largely overrun those due to thrombosis even in worst case scenarios of frequency and gravity of the vaccine side effects.
\end{abstract}

\maketitle

\section*{}
\textbf{We analyze, in the framework of epidemiological modelling, the stop in the deployment of the AstraZeneca vaccine due to some suspected side effects. Indeed, few dozen suspicious cases of Deep Vein Thrombosis (DVT) over 5 millions vaccinations have arisen in Europe and pushed several European countries to suspend AstraZeneca injection. Using both an epidemiological Susceptible-Exposed-Infected-Recovered (SEIR) model and statistical analysis of publicly available data, we estimate the excess deaths resulting from missing inoculations of the vaccine and those potentially linked to DVT side effects in France and Italy. We find that, despite the many simplifications and limitations in our analysis, the excess deaths differ by at least an order of magnitude in the two strategies, that the relative benefits are wider in situations where the reproduction number is larger, and they increase with the temporal duration of the vaccine ban.}

\section{Introduction}

As of March 2021, the spread of the SARS-CoV-2 virus~\citep{WU20} has caused more than 120 millions infections worldwide with a total death toll of more than 2 millions. Up to the end of 2020, the only effective measures to contain the spread of the virus were based on social distancing, wearing face masks and more/less stringent lockdown~\citep{anderson20,Chinazzi20,Yuan20}. Later on, a massive vaccination campaign kicked off in several countries thanks to the availability of a variety of vaccines (e.g., AstraZeneca, Johnson\&Johnson, Moderna, Pfizer/BionTech, Sputnik V). Such vaccines differ substantially in terms of efficacy, legal status, availability, and logistics needed for their delivery to patients. According to various estimates~\citep{Sridhar21}, vaccinations would produce a reduction in infections, and eventually yield to "herd immunity" when $\approx70\%$ of the population gets fully vaccinated. When such a large fraction of the population becomes immune to the disease, its spread from person to person becomes very unlikely, and the whole community becomes protected. By allowing for an earlier easing of non-medical measures against the SARS-CoV-2 virus, vaccination is also expected to significantly reduce the economical, social and psychological impacts of lockdown measures~\citep{Fernandes20}. Those estimates assume that there is no break in the supply of vaccines or any other suspension in the procedure due to side effects from vaccination. Unfortunately, on March 15th 2021 several European countries suspended the use of AstraZeneca COVID-19 vaccine as a precaution in order to investigate the death of a few dozens of patients developing blood clots - associated with Deep Vein Thrombosis (DVT)~\cite{thrombosis2007} - after such  vaccine, despite no proof has been found yet of causal link between vaccination and DVT~\footnote{Bloomberg, March 2021: \texttt{https://tinyurl.com/2zk29abr}}.  
Health personnel who inoculated the vaccine to those who died as a result of DVT are being investigated in Italy for manslaughter~\footnote{ANSA, March 2021: \texttt{https://tinyurl.com/4ywxt5kp}}. 
The contingent situation with the widespread COVID-19 pandemic naturally raises the question of whether a prolonged stop in vaccinations coming from adopting the precautionary principle~\citep{Steele20060} could cause an excess mortality beyond that caused by hypothetical side effects of the vaccines. The European Medicines Agency (EMA) is currently assessing whether the vaccine can continue to be used despite possibly causing this very rare side effect. 
In this Letter, we aim at exploring this issue by computing future COVID-19 epidemic scenarios by comparing i) the excess mortality caused by reducing the vaccinations using the stochastic Susceptible-Exposed-Infected-Recovered (SEIR) model~\citep{Faranda20b}, and ii) the estimates of the possible casualties caused by side effects of a vaccine, namely those associated with DVT. We remark that the additional, longer-term effect of the presence of higher infection rates, e.g. the increased risk of virus mutations leading to possibly more malignant and/or more infectious variants, is not included in our treatment. Our analysis focuses on France and Italy, which have been among the countries that have been most severely impacted by the COVID-19 pandemic~\citep{Alberti20}. An important remark follows. Our goal is not to provide an exact estimate of both i) and ii) but rather to perform an order-of-magnitude comparison between excess deaths resulting from different scenarios of vaccination policy. We proceed in the spirit of complexity science, where simple models are useful for elucidating the main mechanisms behind complex behaviour and provide useful inputs for the deployment of more advanced modelling suites and data collection strategies \cite{Held2005,Pascual2011,Gahde2013,Almaraz2014,GhilLucarini2020}. In other words, we will approach the problem by performing Fermi estimates~\citep{Fermi} where the classical back-of-the-envelope calculations are performed via the SEIR model, allowing to take into account the uncertainties in both model parameters and data. \textit{In nuce}, we perform a counterfactual analysis based on a story-line approach, which has become a powerful investigation method for assessing risks coming from extreme events \cite{counterfactualshepherd}. While the quantitative consolidation of our results clearly requires extensive data analysis and modelling, our findings show with a large confidence that excess deaths due to the interruption of the vaccination campaign largely override those due to DVT even in the worst case scenarios of frequency and gravity of the vaccine side effects. Fermi estimates can provide valuable inputs for an efficient and pragmatic application of the precautionary principle able to reduce the negative impacts of hazards of various nature, as done in economics~\cite{anderson2010applying}. 
\section{Methods}
The model~\citep{Faranda20} with time-dependent control parameters can mimic the dependence on additional/external factors such as variability in the detected cases, different physiological response to the virus, release or reinforcement of distancing measures \citep{Faranda20b}.  Our compartmental model \citep{Brauer08} divides the population into four groups, namely Susceptible (S), Exposed (E), Infected (I), and Recovered (R) individuals, according to the following evolution equations:

\begin{align}
     S_{t+1} & = S_t -\lambda\left(1-\alpha\right)\dfrac{I_{t}S_{t}}{N_{t}}-\lambda\alpha\left(1-\sigma\right)\dfrac{I_{t}S_{t}}{N_{t}}\nonumber\\
     &+\left(1-\sigma\alpha\right)S_{t}\label{stn+1}\\
     E_{t+1} &= E_t + \lambda\left(1-\alpha\right)\dfrac{I_{t}S_{t}}{N_{t}}+\lambda\alpha\left(1-\sigma\right)\dfrac{I_{t}S_{t}}{N_{t}}\nonumber\\
    & +\left(1-\epsilon\right)E_{t}\\
     I_{t+1} &= I_t + \epsilon E_{t}+(1-\alpha-\beta)I_{t}\\
     R_{t+1} &= R_t + \sigma\alpha S_{t} +\beta I_{t}
\end{align}

In the SEIR model above, the classical parameters are the recovery rate ($\beta$), the inverse of the incubation period ($\epsilon$), and the infection rate ($\lambda$). Here we have generalized the model presented in~\citet{Faranda20b} by introducing two additional parameters able to succinctly mimic the strategies of a  vaccination campaign, namely the vaccination rate per capita $\alpha$ and the vaccine efficacy $\sigma$, see ~\citet{sun2010global}. In order to consider uncertainties in long-term extrapolations and time-dependent control parameters, a stochastic approach is used through which the control parameters $\kappa \in \{\alpha, \beta, \epsilon, \lambda, \sigma\}$ are described by an Ornstein-Uhlenbeck process \cite{OU1930} with drift as follows:

\begin{equation}
{\mathrm{d}\kappa} =  -\kappa(t)\mathrm{d}t +\kappa_0\mathrm{d}t +\varsigma_\kappa dW_t,
\label{eqstoc}
\end{equation}

where $\kappa_0 \in \{\alpha_0, \beta_0, \epsilon_0, \lambda_0, \sigma_0\}$, $dW_t$ is the increment of a Wiener process. We remind that the basic reproduction number~\citep{Delamater19} is written as $R_0=\beta_0/\lambda_0$. In Eqs. (\ref{stn+1})-(\ref{eqstoc}) we set $dt=1$, which is the highest time resolution available for official COVID-19--related counts and is relatively small compared to the characteristic times associated with COVID-19 infection, incubation, and recovery/death. 

\begin{table}[h]
    \centering
    \begin{tabular}{|c|c|c|c|c|}
    \hline
        $\alpha_0$  & $\beta_0$  &  $\epsilon_0$  & $\sigma_0$  &   $m_0$  \\ 
        0.0015~\citep[see Ref.][]{ratevaccine}     & 0.37~\citep[see Ref.][]{Faranda20b}      &  0.27~\citep[see Ref.][]{lauer2020incubation}          & 0.59~\citep[see Ref.][]{Astra}       & 0.015~\citep[see Ref.][]{Fanelli20}         \\ \hline
        $\varsigma_\alpha$  & $\varsigma_\beta$  &  $\varsigma_\epsilon$  & $\varsigma_\sigma$  &   $\varsigma_m$  \\ 
        0.25~\citep[see Ref.][]{ratevaccine}      & 0.2~\citep[see Ref.][]{Faranda20}      &  0.2~\citep[see Ref.][]{ratevaccine}          & 0.1~\citep[see Ref.][]{Astra}        & 0.0        \\ \hline
    \end{tabular}
    \caption{Model parameters used for our simulations with corresponding references.}
    \label{tab:parameters}
\end{table}

Initialising parameters with their associated reference are shown in Table \ref{tab:parameters}.
The mortality rate $m_0$ is also shown, set to $0.015$~\citep{Fanelli20}.  While $\beta_0$ and $\epsilon_0$ and the associated $\varsigma$ are the same as in~\citep{Faranda20b}, the values of $\sigma_0$ and respective $\varsigma$ are derived from the range given for the AstraZeneca vaccine phase 3 tests for the first dose~\citep{Astra}, and $\alpha_0$ and $\varsigma_\alpha$ are given supposing that both Italy and France keep vaccinating $10^5$ individuals per day with a 20\% daily fluctuation~\citep{ratevaccine}. As in~\cite{Faranda20b}, we also set $\varsigma_\lambda=0.2$, allowing for 20\% daily fluctuations in the infection rate. Note that here we restrict to Gaussian fluctuations: as shown in~\cite{Faranda20b}, allowing for log-normal fluctuations of the parameters does not change the average results but slightly enhance their dispersion. 
See Supplementary Material for the numerical code. 

\section{Estimate of the Excess Deaths due to Stopping AstraZeneca Vaccine Inoculation}
\begin{figure}[h]
    \centering
    \includegraphics[width=12cm]{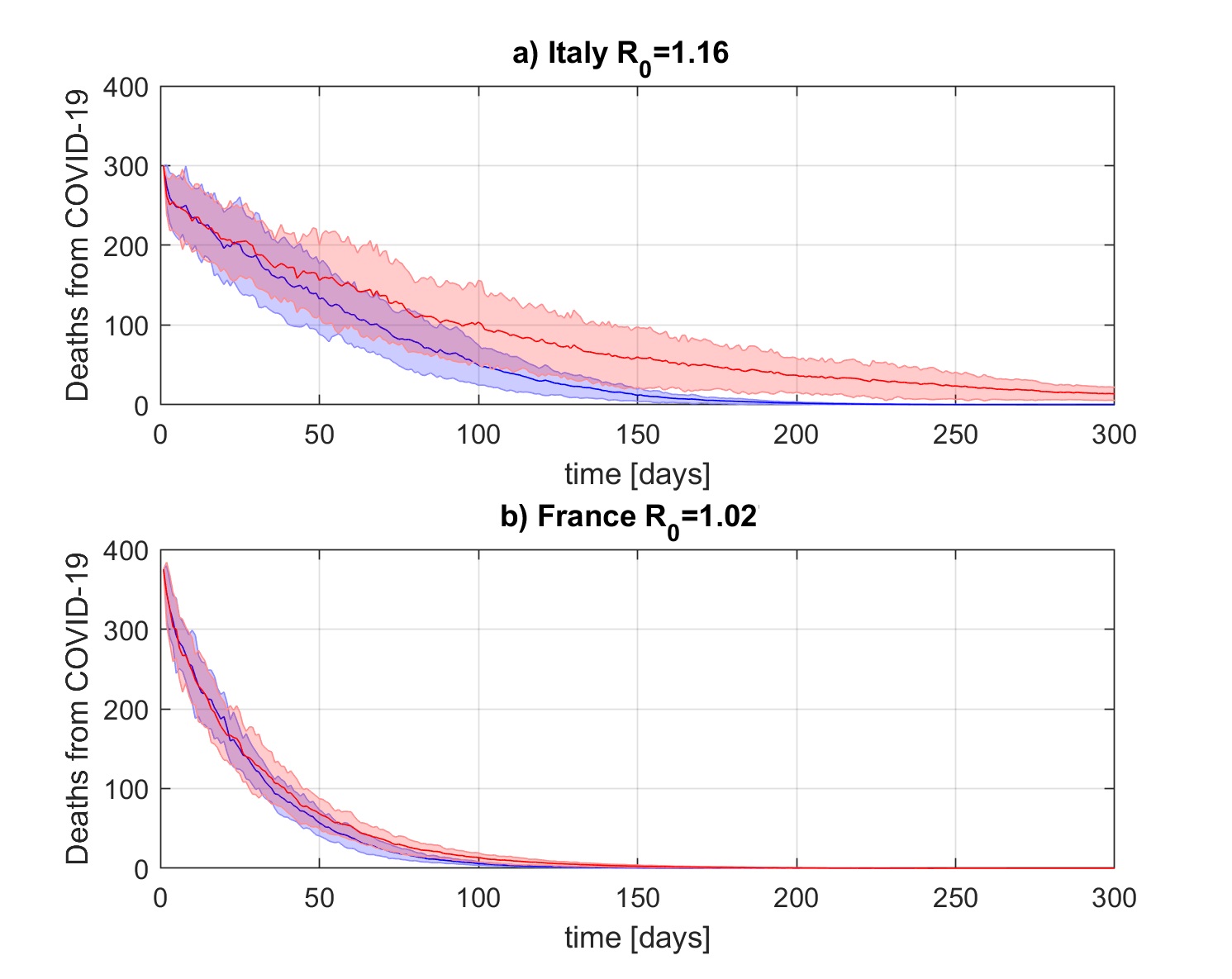}
    \caption{The number of daily deaths $m \times I(t)$ as a function of time (300 out of 500 days shown) for Italy (a) and France (b) using the values of $R_0=1.16$ (Italy) and $R_0=1.02$ reported respectively for the 15th of March countries. Solid lines show the ensemble average, dotted lines extend to one standard deviation of the mean. Red and blue curves refer respectively to no vaccination and a vaccination campaign whose efficacy is 59\%.}
    \label{fig1}
\end{figure}

Figure \ref{fig1} reports the daily number of deaths $m_0 \times I_{t}$ as a function of time for Italy (a) and France (b). Initial conditions are set for both countries to the values reported on March 15th as follows: for Italy, we set $N=60\cdot10^6$ population, $E_{t=1}=I_{t=1}=20\cdot10^4$ as the infected and exposed populations, $R_{t=1}=11\cdot10^6$ as the sum of $9\cdot10^6$ recovered estimated from serologic tests and $2\cdot 10^6$ immunized from 2 doses of either Pfizer/BioNTech, Moderna or AstraZeneca vaccines and $R_0=1.16$.  For France, we set $N=67\cdot10^6$, $E_{t=1}=I_{t=1}=25\cdot10^4$, $R_{t=1}=13.2\cdot10^6$ as the sum of $11\cdot10^6$ recovered estimated from serologic tests and $2.2\cdot 10^6$ immunized from vaccines and $R_0=1.02$. For both France and Italy, we assume that the virus, after the second wave, has infected the 15\% of the population. This estimates are based on~\citet{pullano2021underdetection} who reported a 7\%$\pm$ 3\% total infections for France after the first wave, assuming that the second wave had a similar magnitude for both countries. We remark however, that our results are basically insensitive to oscillation of $S(1)$ of order of 5 millions individuals (cfr. Supplementary Material Figure S1). Rather than integrating the Fokker-Planck equation \citep{Risken1996} corresponding to the system of equations given above, we follow a Monte Carlo approach and we perform two sets of $N_r=1000$ realizations (see supplementary material Figure S2 for a justification of this value): stopping (red) and continuing (blue) the vaccination campaign at the same rate. The model is integrated for 500 days, that is about the time it would take to vaccine the rest of the susceptible population with AstraZeneca at the rate of 10$^5$ individuals per day.

First, we observe a monotonic decrease in the daily deaths for all scenarios considered from the initial date $t=t_0$ corresponding to March 15, 2021. This is in agreement with actual estimates that for Italy and France the so-called third wave should reach its peak in the second half of March, 2021~\footnote{The Guardian, March 2021: \texttt{https://tinyurl.com/vvwhcydz}}
Moreover, we observe that the cumulative number of deaths significantly (we take the width of the error bars as level of significance) reduces if vaccinations are continued at 100000 doses per day with respect to the scenario where vaccination is stopped. For Italy (France) completely halting the vaccination, at the actual epidemic rate, the number of excess deaths from COVID19 would amount to $9\pm 3 \cdot 10^3$ ($1.2 \pm 0.4 \cdot 10^3$) excess deaths from COVID19.
The difference between the two countries is largely due to the value of $R_0$, which is larger for Italy. This suggests that halting vaccination in a growing epidemics phase (Italy) has more dramatic consequences than in a more controlled scenario of $R_0\approx1$ (France). 
  
\begin{figure}[h]
    \centering
    \includegraphics[width=14cm]{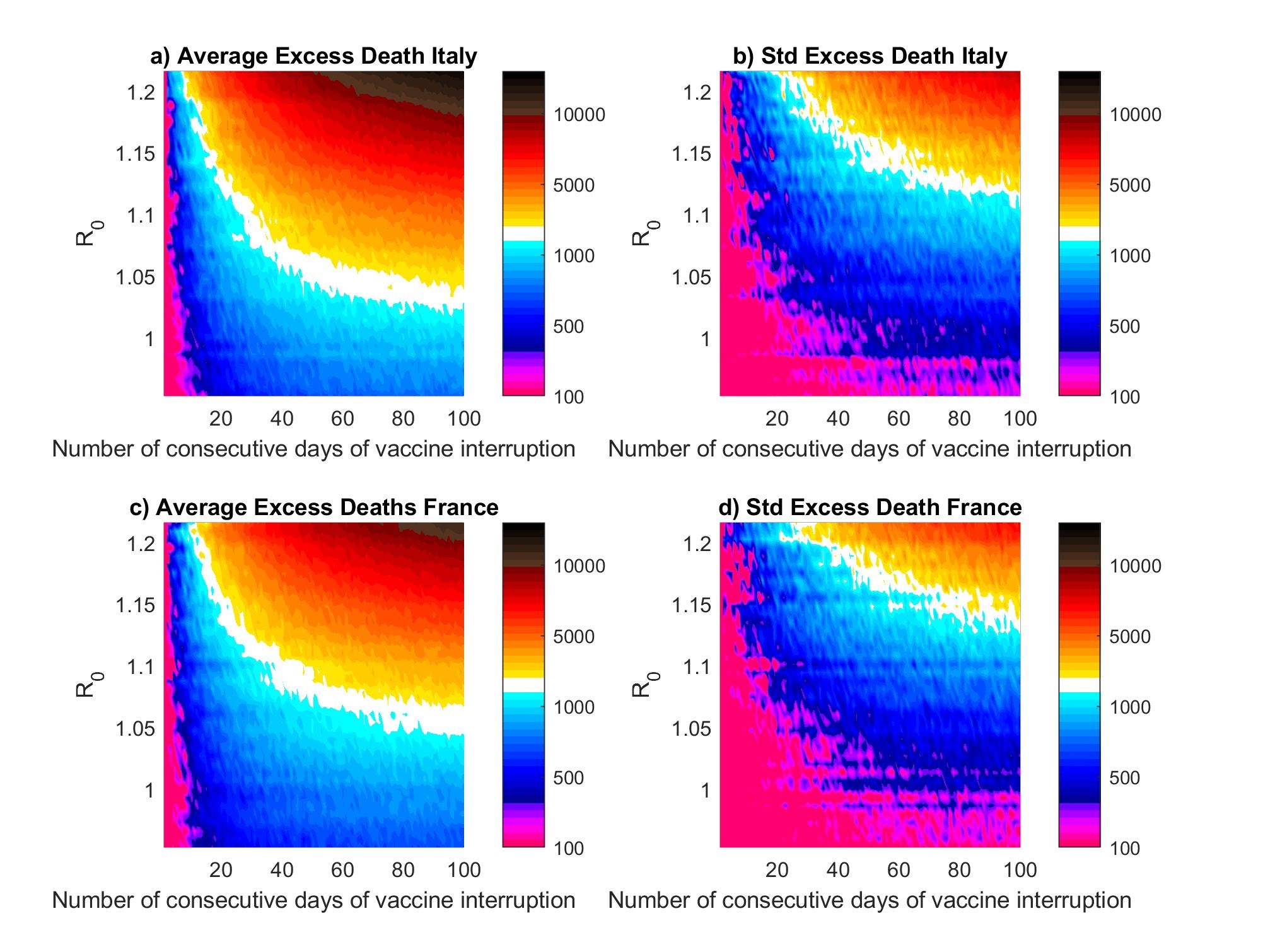}
    \caption{(a,c) Average and (b,d) standard deviation over $N_r = 1000$ realizations of the stochastic SEIR model showing the  excess deaths $m \times I(t)$ as a function of the number of the days of interruption of AstraZeneca vaccinations (x-axis) and $R_0$ (y-axis) for Italy (a,b) and France (c,d). The excess deaths are computed with respect to a base scenario where vaccine injections are never interrupted. Note that $x$-axis starts at $N=1$. Each realization of the SEIR model is integrated for 500 days.}
    \label{fig2}
\end{figure}

\begin{figure}[h]
    \centering
    \includegraphics[width=12cm]{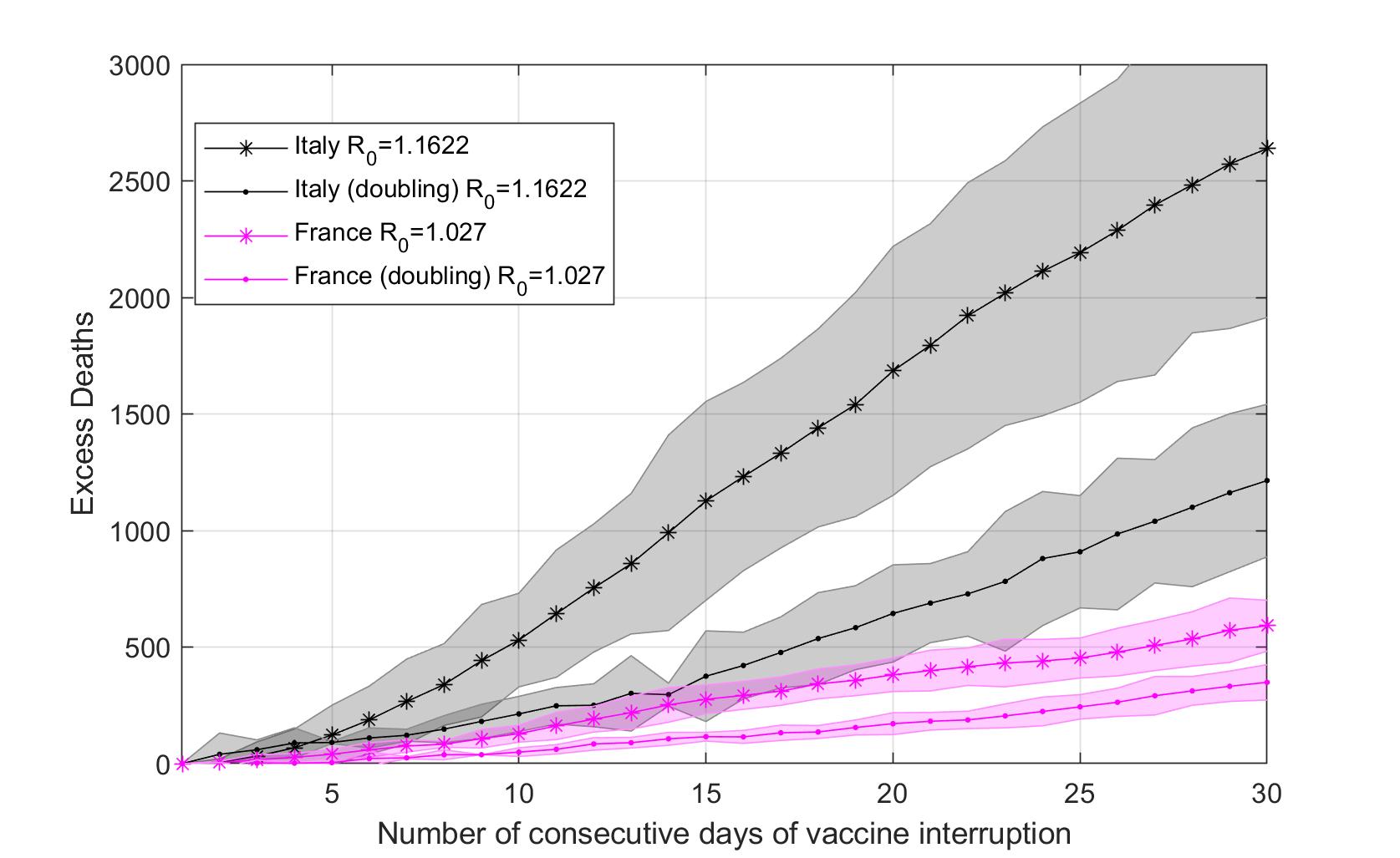}
    \caption{Average over $N_r = 1000$ realizations of the stochastic SEIR model showing the  excess deaths $m \times I(t)$ as a function of the number of the days of interruption of AstraZeneca vaccinations for Italy $R_0=1.16$ (black and France $R_0=1.02$ (magenta). Simulations are smoothed with a moving average filter with window size 10 days. Stars indicate simulations where vaccination are resumed at the same rate, dots simulations where the vaccination rate is doubled for a number of days equivalent to those of interruption.  Error bars are computed as the mean relative error. }
    \label{fig3}
\end{figure}

Our previous analysis is based on a total stop of AstraZeneca vaccination. However, a more realistic scenario is to assume that AstraZeneca vaccination will resume after a limited number of days used for verification. We investigate this effect in Fig.~\ref{fig2}. There, we consider the average excess deaths as a function of the interruption length in number of days (x-axis) and $R_0$ (y-axis) for Italy (a) and France (b). The excess deaths are computed with respect to a base scenario where vaccine injections are never interrupted and they are averaged over 1000 realizations of the SEIR model. Figure~\ref{fig2} shows that the longer is the vaccine injections disruption, the higher is the number of excess  deaths. The impact is stronger for higher values of $R_0$. While waiting the advice of EMA about AstraZeneca safety, many national health agencies also announced that, when allowed, they would resume the vaccination at a higher rate than before to override the effects of the stop. In the supplementary Figure S3 we therefore present a set of simulation where, for a number of days equal to those of the vaccination interruption, injections are performed at a double rate than originally planned, i.e., 2$\cdot 10^5$ individuals/day, in order to compensate for the lost vaccinations. Although reduced, the number of excess deaths is still high and of the same order of magnitude as the one estimated in Fig.~\ref{fig2}, as a result of the nonlinear cascade effect of the extra infections occurred in the period when vaccinations were interrupted. A focus on the actual values of $R_0$ for Italy and France is reported in Fig.~\ref{fig3}. Here we compare the two countries and we also show the effect of doubling vaccination rates. This shows that excess deaths scale down by a factor two but they remain of the same order of magnitude as for the case of a business-as-usual vaccination rate, namely $10^5$ vaccinations/day. 

\section{Worst case scenarios for AstraZeneca Side Effects}
The final step in our investigation is to compare the previous estimates of excess deaths with an order of magnitude estimate of deaths due to DVT resulting from side effects of the AstraZeneca vaccine. In order to make a meaningful comparison, in a case where uncertainties are very large and hard to quantify, we will consider a worst case scenario for the impacts of the side effects. This scenario relies on the unrealistic hypothesis that the totality of susceptible population to DVT suffers from DVT shortly after being vaccinated, and the lethality rate is similar to the one observed in the overall population. 

As of March 15th 2021, few dozens suspect cases of DVT have been reported over a number of 5 millions vaccinated people with AstraZeneca in Europe\footnote{here, we use European data accessible via the website of the European Medicines Agency at \texttt{https://tinyurl.com/ht8y98kr} 
to average out the large spread of national data.}. By suspect cases we mean people who have developed DVT in the few days following the vaccination. This leads us to an estimate of a frequency of 6 cases per million of vaccines. Let us call this rate $r_{AZ}^{DVT}$. Let us also consider that, in the case of France, the incidence of DVT has been estimated to 1800 people per 1 million inhabitants per year (\cite{Bouee2016}), with a lethality rate after three months of 5\% \cite{Heit2015}, raising to 30\% when a period of 5 years is considered \cite{thrombosis2007}. This leads to estimating a total of the order of 10000 deaths per year as a result of DVT. Even assuming that all DVT cases following the inoculation of the AstraZeneca vaccine would have not manifested themselves in absence of the injection, we have that $N$ vaccinations would lead to an extra $N \times r_{AZ}^{DVT}$ DVT cases. Let us assume that all of these cases result into death\footnote{Current data suggest that this is manifestly a gross worst case approximation.}. We then have that $10^5$ daily vaccinations would result into a maximum of 0.6 daily deaths. In 500 days, which is the time needed to cover the entirety of the French population, this leads to an upper bound of 300 deaths. Considering a death rate of $30\%$, the number scales down to approximately 100, while considering a death rate of $5\%$ the number scales down to approximately 15. Similar figures apply for Italy. 

\section{Conclusion}
Decision-making in presence of strong uncertainties associated with health and environmental risks is an extremely complex process, resulting from the interplay between science, politics, stakeholders, activists, lobbies, media, and society at large \cite{AGU2010,Benessia2017,Reis2019}. In this letter, we have aimed at contributing to the debate on different strategies for combating, in conditions of great uncertainties in terms of health and social response, pandemic like the current one caused by the SARS-CoV-2 virus. We have focused on the case of the AstraZeneca COVID-19 vaccine and on the locales of Italy and France, for the period starting on March 15th 2021. The goal is providing a semi-quantitative comparison, based on Fermi estimates informed by a simple yet robust stochastic model, between the excess deaths due to temporal restriction in the deployment of a still experimental vaccine and the excess deaths due to its possible side effects. Given the many uncertainties on the (possible) side effects of the vaccine, we have resorted to making worst case scenario calculations in order to provide a robust upper bound to the related excess deaths. Our results are preliminary and should be supplemented by more detailed modelling and data collection exercises. Indeed: i) we assume a single vaccine with the nominal AstraZeneca efficacy, neglecting the other available vaccines, ii) we consider a fixed vaccination rate, iii) for AstraZeneca DVT side effects we consider French data and rescale them for the Italian populations, iv) we focused our analysis on DVT side effects, but other pathologies could be considered with the same approach. Yet, these results clearly suggest - see a useful summary in Table \ref{tab:vacc} - that the benefits of deploying the vaccine greatly outweigh the associated risks, and that the relative benefits are wider in situations where the reproduction number is larger, and they increase with the temporal duration of the vaccine ban. We have also analysed the case of resuming the vaccinations at a double rate ($2\cdot 10^5$ vaccinations/day) for an amount of days equal to vaccine interruption period (Fig.~\ref{fig3} and Fig. S3). This analysis has pointed out that excess deaths are still of the same order of magnitude as those observed by resuming vaccinations with $10^5$ vaccinations/day injection rate but scale down by a factor 2. This is an evident outcome of the nonlinear effects of epidemiological dynamics: those who have not been vaccinated can contaminate other individuals before vaccination resume, as a result of a cascade mechanism also observed in turbulent flows: there,  energy injected in large scales vortex is transferred to small scales via nonlinear interactions between scales \citep{Kolmogorov41}. Here, in analogy, a few non-vaccinated individuals can produce a large number of infected individuals. The process can only stop if a huge number of daily vaccinations (much larger than a factor 2) is performed. Nevertheless, this still requires a characteristic recovery timescale $T$ that is larger than the typical immunization scale $\eta$ (e.g., a few months for AstraZeneca~\citep{Astra}). Finally, even if several countries have resumed, or are going to resume, AstraZeneca vaccinations, the effect of the interruption is hard to counterbalance and require vaccination efforts difficult to set-up in due times. Furthermore, at least for large countries where AstraZeneca vaccination could resume, the confidence of the population in the vaccines is reduced by a non negligible percentage \footnote{The Economist, March 2021: \texttt{https://tinyurl.com/83cbr4d3}}.
In this sense, our estimates are likely to be conservative and might possibly underestimate the excess deaths deriving from the disbelief in the vaccination policies observed in the largest European countries. The analysis presented here has been performed with a parsimonious but well-posed and tested model and we hope that the results we obtain might be the starting point for more detailed, more advanced, and more mature investigations with sophisticated models and data collection exercises.

\begin{table}[]
    \centering
    \begin{tabular}{|c|c|c|}
    \hline
        Excess Deaths & Italy & France  \\ \hline
        Stop AZ for $t=500$ days & $ 9000\pm 3000$  & $1200 \pm 400$ \\ \hline
        Stop AZ for $t=14$ days & $1700\pm 500$   & $430 \pm 70$ \\ \hline
        Stop AZ for $t=7$ days & $ 790\pm 90$  & $160 \pm 30$ \\ \hline
        Stop AZ for $t=3$ days & $ 260\pm 50$ & $130 \pm 20$ \\ \hline
        Worst case DVT deaths due to AZ & $\approx280$ &  $\approx300$ \\ \hline
        High fatality DVT deaths due to AZ & $\approx90$ &  $\approx100$ \\ \hline
        Standard fatality DVT deaths due to AZ & $\approx13$ &  $\approx15$ \\ \hline
    \end{tabular}
    \caption{The first 4 lines of the table indicate the excess deaths due to the interruption of AstraZeneca  compared to a reference scenario where the vaccine injections are never interrupted. The SEIR model is integrated for 500 days with $R_0$ = 1.16 for Italy and $R_0$ = 1.02 for France. The last 3 rows of the table show the deaths from deep vein thrombosis (DVT) that could be due to the vaccine in three different scenarios: the worst case (100\% mortality rate), a high mortality scenario (death rate of 30\%) and a standard mortality scenario (5\% mortality rate) assuming a period of 500 days.}
    \label{tab:vacc}
\end{table}

\section*{Acknowledgements}
This work has been greatly supported by the London Mathematical Laboratory and we acknowledge the logistic support of SCuP.  VLu acknowledges the support received from the EPSRC project EP/T018178/1 and from the EU Horizon 2020 project TiPES (Grant no. 820970). This is TiPES' contribution \#97. We thank A Veber, A Mazaud, FM Breon, the Modcov19 CNRS community and two anonymous reviewers for useful comments and suggestions.

\section*{Data Availability}

Raw data that support the findings of this study are openly
available in Johns Hopkins University Center for Systems Science at
\texttt{https://systems.jhu.edu/research/public-health/ncov/}. Derived data
supporting the findings of this study are available from the corresponding
author upon reasonable request.

\end{document}